\RequirePackage{fix-cm}
\documentclass[smallextended]{svjour3}       
\smartqed  
\usepackage{graphicx}
\usepackage{latexsym}
\begin{document}

\title{Stealth's cosmologies}


\author{Abigail Alvarez                     \and
             Cuauhtemoc Campuzano      \and
             V\'ictor C\'ardenas                  \and
             Efra\'in Rojas 
}



\institute{Abigail Alvarez \at
           Facultad de F\'\i sica, Universidad Veracruzana,
           91000, Xalapa, VER., M\'exico \\
           \email{abialvarez@uv.mx} \\
           \and
           Cuauhtemoc Campuzano \at
           Facultad de F\'\i sica, Universidad Veracruzana,
           91000, Xalapa, VER., M\'exico \\
           \email{ccampuzano@uv.mx} \\
           \and
           V\'\i ctor C\'ardenas \at
           Instituto de F\'{\i}sica y Astronom\'ia, Universidad de Valpara\'iso, 
           Gran Breta\~na 1111, Valpara\'iso, Chile \\
           \email{victor.cardenas@uv.cl} \\
           \and
           Efra\'\i n Rojas  \at
           Facultad de F\'\i sica, Universidad Veracruzana,
           91000, Xalapa, VER., M\'exico \\
           \email{efrojas@uv.mx} \\     
           }

\date{Received: date / Accepted: date}

\maketitle

\begin{abstract}
We present a novel approach to construct cosmological models endowed with a particular scalar field, 
the stealth. The model is constructed by studying a scalar field no-minimally coupled to
 the gravitational field with sources; as sources, we used a perfect fluid and analyzed the simplest 
 case of dust and the power-law cosmology. Surprisingly, we find that these stealth fields, which have 
 no back-reaction to background space-time, have a contribution to cosmological dynamics. 
 Furthermore, we provide analytic expressions of the stealth's contributions to the energy density in 
 both cases, and for the pressure in the power-low cosmology, which means that such contributions 
 to cosmological evolution are quantifiable. Additionally, we discuss the behaviors of the 
 self-interaction potential for some cases.
\end{abstract}

\section{Introduction}

\label{intro}
The stealth fields are nontrivial solutions to scalar fields non-minimally coupled to the gravitational 
field with a vanishing stress-energy tensor. This particular fact does not induce a back-reaction on 
the background spacetime where the stealths propagate. The stealth fields based on the Einstein 
field equations as well as in some modified gravity field equations have been fairly studied in the 
last few years where their existence has been proved in several gravitational contexts 
\cite{rubens,poloco,AyonBeato:2005tuA,AyonBeato:2004ig,banerj,mokthar,mokthar2,mokthar3,faraoni,mokthar5,mokthar6,cisterna,charmois,temoc,Smolic,chilenos,maeda,Ayon-Beato:2013bsa,Ayon-Beato:2015bsa}. 
In any case, despite the wealth of references quoted the 
subject, from the field equations little can be concluded regarding the interaction between a stealth 
field and ordinary matter. In this sense, we intend to develop further such subject in the present 
work. In fact, remarkable results in this direction were provided in \cite{poloco} where a deep 
analysis of the interaction of matter which does not carry energy and the ordinary matter, 
was performed.

One of the most outstanding conclusions from the current astronomical data has been the 
confirmation of the late cosmic acceleration of the Universe \cite{SnIa}. Until now, the best fit to 
this data is by using the well known $\Lambda$ Cold Dark Matter ($\Lambda$CDM) model. This 
model necessarily includes a cosmological constant to make it work and exhibit an accelerated 
behavior of the Universe but the criticism is that such constant is included by hand into the 
model. This fact leaves several questions and concerns, mainly about the nature of the constant; 
for example, its origin and value that are very close to critical energy density confronting thus the 
predictions made by the quantum field theory in approximately 120 orders of magnitude. There 
are two main approaches pursued to describe the phenomenon of the cosmic acceleration: the 
first one is based in the inclusion of an exotic matter, i.e. the dark energy (DE) where the 
cosmological constant drives the expansion; the second one, by modifying the Einstein gravity 
theory where some non-standard geometrical terms lead to an accelerated performance. While 
these models are interesting, it is probably fair to say that their results about the accelerated 
behavior of the Universe are not conclusive, especially when they are compared to those
 from $\Lambda$CDM model.

On the other hand, the stealth fields in the cosmological context have been a focus of great interest 
since the so-called dark energy sector of the Universe seems to be a special arena where modified 
gravity theories, and observations, may be contrasted. In this spirit, in \cite{1972} was shown that 
the presence of a stealth field in a cosmological scenario is able to reproduce the observational data. 
In view of this situation, encouraged by their particular behaviors as well the possibility to grasp 
their role into the cosmological setup, one may wonder if the stealth field evolution in a 
Friedmann-Lema\^{\i}tre- Robertson-Walker (FLRW) spacetime coupled to a perfect fluid indicates 
a viable alternative to explain the accelerated expansion of the universe. There are other approaches 
that pursue the same goal such as quintessence \cite{shani}, Chaplygin gas \cite{chaplin} and its 
generalization \cite{chaplinG}, $k$-essence \cite{kessence}, to mention something. In all these 
cases, they invoke a scalar field to drive the cosmic acceleration. Following this line of reasoning, 
in this paper, we construct a cosmological model endowed with a scalar field, the stealth field, 
with the particular fact that it does not back-react on an FLRW background spacetime, contrary to 
the above mentioned models. As a result, we obtain a general approach relating the stealth with 
any type of matter coupled to the gravitational field.

In this work, we attempt to carry one step further the FLRW geometry as a background spacetime 
where the stealth field $\phi$ propagate, endowing it with matter content provided by dust. By 
respecting the vanishing of the stress tensor associated to the stealth as field equations, we 
explicitly construct solutions for $\phi$ as well as the self-interacting potential. At dependence of 
the parameters of the theory, this may (or not) have some non-trivial impacts on the cosmology 
of the Universe.

The paper is organized as follows. In section 2 the general approach for the stealth in presence of 
sources is given. In section 3 we used the approach to study the stealth in the presence of dust. 
In section 4 an example of the stealth in a power law universe expansion is studied, finally, in 
section 5 we end with a discussion and conclusions are given.
   

\section{Stealths in presence of sources}
\label{sec:1}

The stealth fields are characteristics of Hordensky theories of gravitation where the 
scalar fields are coupled non-minimally to gravity. In the present work, we study the stealth in an 
FLRW space-time with a perfect fluid and its cosmological consequences. An action describing such 
a scenario is,

\begin{eqnarray}\label{eq:action}
S[g_{\mu\nu},\phi]=\int d^4x \sqrt{-g} \left[ \frac{R}{2\kappa} + L_m - \frac{1}{2} \zeta R \phi^2
- \frac{1}{2} \partial_{\mu} \phi \partial^{\mu} \phi
- V(\phi) \right].
\end{eqnarray}
In the above expression, $\zeta$ is the coupling constant, $L_m$ is the matter
Lagrangian, $\phi$ is the scalar field, and $V(\phi)$ is the self-interaction potential of the scalar field.

The field equations obtained from the variation of the action (\ref{eq:action}) with respect to the 
metric are written as
\begin{eqnarray}\label{eq:feq}
 G_{\mu\nu}-\kappa T^{(m)}_{\mu\nu}= \kappa T^{(S)}_{\mu\nu},
\end{eqnarray}
where $T^{(m)}_{\mu\nu}$ is the stress-energy tensor of ordinary matter and
$T^{(S)}_{\mu\nu}$ is the stress-energy tensor of the stealth field
$\phi$; explicitly:
\begin{equation}\label{eq:stealtuv}
T^{(S)}_{\mu\nu}=\nabla_\mu \phi \nabla_\nu \phi -( V(\phi)+
   \frac{1}{2}\nabla_\alpha \phi \nabla^\alpha \phi ) g_{\mu\nu}
  +
  \zeta (G_{\mu\nu}-\nabla_\mu \nabla_\nu
  +g_{\mu\nu} \nabla^\alpha \nabla_\alpha  )\phi^2. 
\end{equation}
Meanwhile, from variation with respect to the scalar field, the equation which describes its dynamics 
is obtained 
\begin{eqnarray}\label{eq:KGeq}
\nabla^\mu\nabla_\mu \phi=\zeta R\phi +\frac{dV(\phi)}{d\phi}.
\end{eqnarray}

The existence of a stealth field for a given background is established by the vanishing of  (\ref{eq:feq});
\begin{equation}\label{eq:stcond}
T_{\mu\nu}^{(S)}=0,
\end{equation}
and its solutions $\phi$ and $V(\phi)$ must satisfy the equation which describes the dynamics of 
scalar field,
\begin{eqnarray}\label{eq:KGeq}
\nabla^\mu\nabla_\mu \phi=\zeta R\phi +\frac{dV(\phi)}{d\phi},
\end{eqnarray}
obtained by the variation of (\ref{eq:action}) with respect to the scalar field. At this point, the problem 
is to solve the system of equations (\ref{eq:stcond}) and (\ref{eq:KGeq}).
However, as was pointed out in \cite{mokthar} the fulfillment of Eq.(\ref{eq:KGeq}) is warranted from 
the conservation of the energy-momentum stealth tensor due to diffeomorphism invariance of the 
action (\ref{eq:action}), 
\begin{equation}\label{eq:divT}
\nabla^\mu T_{\mu \nu}^{(S)}=0,
\end{equation}
which means that solutions of (\ref{eq:stcond}) satisfying (\ref{eq:divT}) are necessarily solutions of 
(\ref{eq:KGeq}). 

In the present work, for the construction of the stress-energy tensor for the stealth field 
$T^{(S)}_{\mu\nu}$, we give the matter information of the Einstein's equations instead of the 
geometrical one, so then the ordinary matter it is related with the stealth field through the 
expression,

\begin{eqnarray}\label{eq:perrona}
  T^{(m)}_{\mu\nu}&=&\frac{1}{\kappa\phi^2}\Big\{-\frac{1}{\zeta}\left[
  \nabla_\mu \phi \nabla_\nu\phi - \left(V(\phi)+\frac{1}{2}\nabla_\alpha
  \phi\nabla^\alpha \phi
  \right)g_{\mu\nu} \right]
  \nonumber\\
 &{}& +\left(\nabla_\mu \nabla_\nu \phi^2
  -g_{\mu\nu} \nabla^\alpha \nabla_\alpha \phi^2\right)\Big\}.
\end{eqnarray}
It is remarkable that this expression is consistent just only when stealth configurations exist.
In what follows, we will consider that scalar field $\phi$ is homogeneous, i.e. depend only on time.
\section{Stealth's cosmological models}\label{sec2}

Bearing in mind the picture of the Universe by the Big Bang theory, we will model it with a perfect 
fluid in an FLRW background. 
\begin{equation}
ds^2=-dt^2+\frac{a^2}{1-kr^2}dr^2+a^2(r^2d\theta^2+r^2 \sin(\theta)^2 d\phi^2),
\end{equation}
as usual, the scale factor $a$ is a function of time. 

As a source we use a perfect fluid stress-energy tensor 
$T_{\mu\nu}^{(m)}=(\rho + p)u_{\mu}u_\nu + p g_{\mu\nu}$, where $p=p(t)$ is the pressure, 
$\rho=\rho(t)$ the energy density and $u^\mu$ is the $4$-velocity of the fluid with 
$u^\mu u_\mu=-1$. By substituting this tensor into (\ref{eq:perrona}) we obtain;
\begin{eqnarray}
\zeta\kappa p(t)^{(GR)} &=& - \zeta\frac{d}{dt}\ln{\phi} \cdot \frac{d}{dt}\ln
\left(a^{-4}\phi^{\frac{1-4\zeta}{2\zeta}}\dot{\phi}^{-2}\right)+\frac{V(\phi)}
{\phi^2}, \label{eq:p1}
\\
\nonumber\\
\zeta\kappa\rho(t)^{(GR)} &=&-\frac{d}{dt}\ln\phi\cdot\frac{d}{dt}\ln(
a^{6\zeta}\phi^{1/2})-\frac{V(\phi)}{\phi^2},
\label{eq:rho1}
\end{eqnarray}
where $\dot{f}:=df/dt$; there are two independent equations, one from the spatial component and 
the other from the time component.

The cosmological model is realized combining Eq.(\ref{eq:rho1}) and using the fact
that the density in that equation must correspond with the given by GR, then it
must satisfied the Friedmann equation $H^2=\kappa\rho/3$, so
\begin{equation}
H^2=-\frac{1}{3\zeta}\left[ \frac{d}{dt}\ln\phi\cdot\frac{d}{dt}\ln(
a^{6\zeta}\phi^{1/2})+\frac{V(\phi)}{\phi^2}\right].
\end{equation}
The simplest case to study is the dust case, that is, $p(t)=0$. So from (\ref{eq:p1}) for the 
self-interacting potential we get
\begin{eqnarray}\label{eq:V1}
V(\phi)  =
\zeta\phi^2\frac{d}{dt}\ln{\phi} \cdot
\frac{d}{dt}\ln
\left(a^{-4}\phi^{\frac{1-4\zeta}{2\zeta}}\dot{\phi}^{-2}\right),
\end{eqnarray}
and for the energy density, from (\ref{eq:rho1}) and (\ref{eq:V1}),
\begin{eqnarray}\label{eq:rho}
 \kappa \rho 
= \frac{d }{dt}\ln \phi\cdot \frac{d}{dt}\ln\left(\frac{ \dot{\phi}
 \phi^{(2\zeta-1)/2\zeta}}{a } \right)^2.
\end{eqnarray}
Now the problem has been reduced to finding the stealth field $\phi$, while the energy
density and the potential are given by the above equations.\\
 
In the dust case, the non-relativistic matter contribution behaves as
$\rho \simeq a^{-3}$. Then, we try to integrate Eq. (\ref{eq:rho}) using the density
$\rho \propto a^{-3}$, however, the direct substitution of the $\rho$ in
(\ref{eq:rho}) leads us to a Riccati equation and for this case, there is no analytic
solution. We are interested in the analytic solutions of Eq.  (\ref{eq:rho});
therefore, we propose to seek solutions for $\phi$ that depends only on the scale
factor, i.e., $\phi=\phi(a)$, so Eq. (\ref{eq:rho}) turns out to be
\begin{eqnarray}
\frac{d}{da}\ln\phi\cdot\frac{d}{da}\ln\left(\frac{\phi^{\frac{2\zeta-1}{2\zeta}
}\phi'}{a^{\frac{3}{2}}}\right)&=&\frac{3}{2}\frac{1}{a^2}\label{eq:F},
\end{eqnarray}
where we have defined $f':=df/da$, (for more detail see appendix \ref{app:A}). It is worth to remark 
that the solutions to the above equation have different behaviors depending on the range of possible 
values of $\zeta$, as we will see below.

\subsection{Homogeneous cosmologies for  $\zeta \in (-\infty , 0) \cup (12/73, \infty)-\{1/4\}$ }

By integrating (\ref{eq:F}), we obtain for the scalar field
\begin{eqnarray}\label{eq:Sol_phi}
  \phi(a)=\Phi_0\left[\left(\frac{a}{a_0} \right)^{5/4+\beta}
  -\left(\frac{a}{a_0}\right)^{5/4-\beta}\right]^{\frac{2\zeta}{4\zeta-1}}
\end{eqnarray}
where
\begin{eqnarray}
\Phi_0=\left(\phi_0^{\frac{5/4+\beta}{2\beta}}
  \phi_1^{\frac{5/4-\beta}{2\beta}}
  \right)^{\frac{2\zeta}{4\zeta-1}},\quad
  a_0=\left(\frac{\phi_0}{\phi_1}\right)^{\frac{1}{2\beta} }\quad \mbox{and}
  \quad \beta=\frac{\sqrt{\zeta(73\zeta-12)} }{4\zeta},\nonumber
\end{eqnarray}
here $\phi_0$ and $\phi_1$ are integration constants.
Now, for the same range of values for $\zeta$, the potential is obtained by substitution of
the above equation into (\ref{eq:V1}),
\begin{eqnarray}
V(a)=\kappa\rho_0\frac{\phi^2}{a^3}\left\{V_2-\left[V_1
  \left(\frac{a}{a_0}\right)^{\beta}+V_0\left(\frac{a}{a_0}\right)^{-\beta}
\right] \right\} \left(\frac{a}{a_0}\right)^{-\frac{1}{2}}\left(\frac{\phi}
{\phi_0}\right)^{\frac{1-2\zeta}{\zeta}}
\end{eqnarray}
where we have defined $V_2:=8(16\beta^2+35)\zeta-48\beta^2-45$ and
$V_{1,0}:=8(4\beta\pm 7)(4\beta\pm 5)-(4\beta\pm 9)(4\beta\pm 5)$. It is verify by substitution 
of (\ref{eq:Sol_phi}) into (\ref{eq:rho}) the right expression to the density $\rho = \rho_0/a(t)^3$.

At this point, we have obtained a stealth model that mimics exactly a matter dominated universe. It is 
important to remark that $\rho$ has to be positive and this would eventually implies a new range for 
the coupling $\zeta$.

In the following figure we show the graphs of $V(\phi)$ vs. $\phi$, for four
different values of the parameters in $V$.

\begin{figure}[h!]
  \begin{center}
\includegraphics[width=0.45\textwidth]{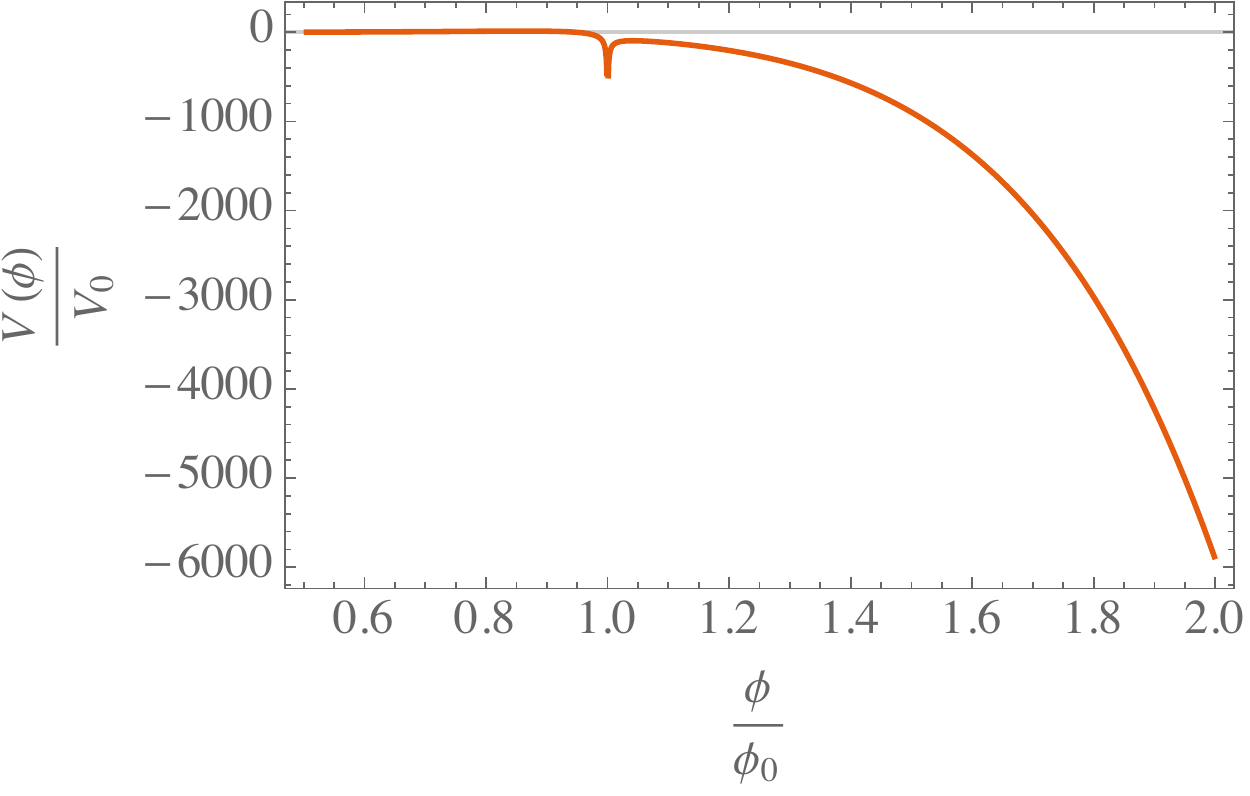}
\includegraphics[width=0.45\textwidth]{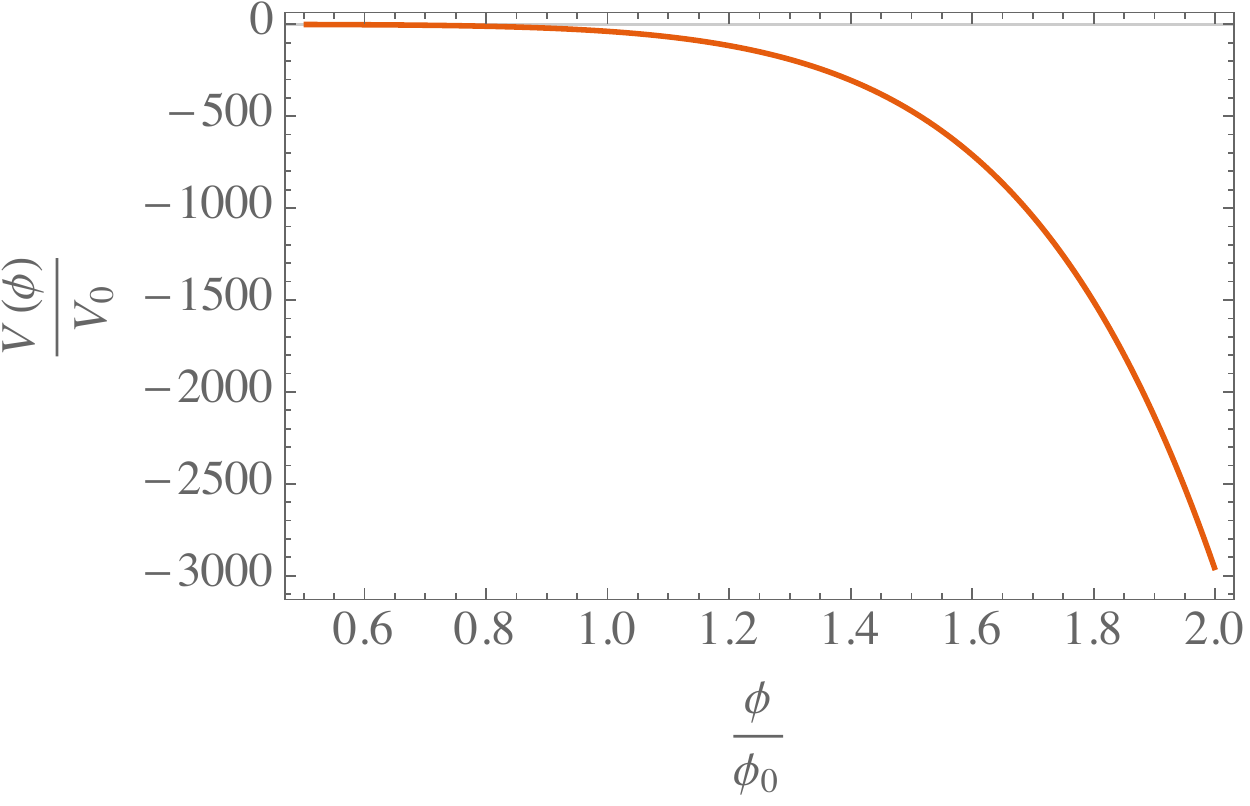}
\caption{The left panel shows $V(\phi)$ vs. $\phi$
for the values $\zeta=1$, $\phi_0=1$, $\phi_1=1$, $\kappa=1$, and $\rho_0=1$; and
at the right panel for $\zeta=1$, $\phi_0=-1$, $\phi_1=1$, $\kappa=1$, and $\rho_0=1$.
\label{fig:fig1}}
\end{center}
\end{figure}
As is clear from Fig.(\ref{fig:fig1}) the potential is nearly flat near $\phi = 0$ and positive as the values 
of $\phi$ increase. The same happens for both cases, although the left panel shows a sharp deep well 
around $\phi = 1$. It is interesting to note that for values of $\zeta<0$, the discriminant of the root 
in $\beta$ always become positive so for this case the solutions given by Eq. (\ref{eq:Sol_phi}) remain 
valid. From the analysis of this solution shows that stealths can appear at the same time as spacetime, 
as we will see later it is not obvious, some solutions allow stealths sometime after the cosmological 
evolution begun. 

\begin{figure}[h!]
  \begin{center}
\includegraphics[width=0.45\textwidth]{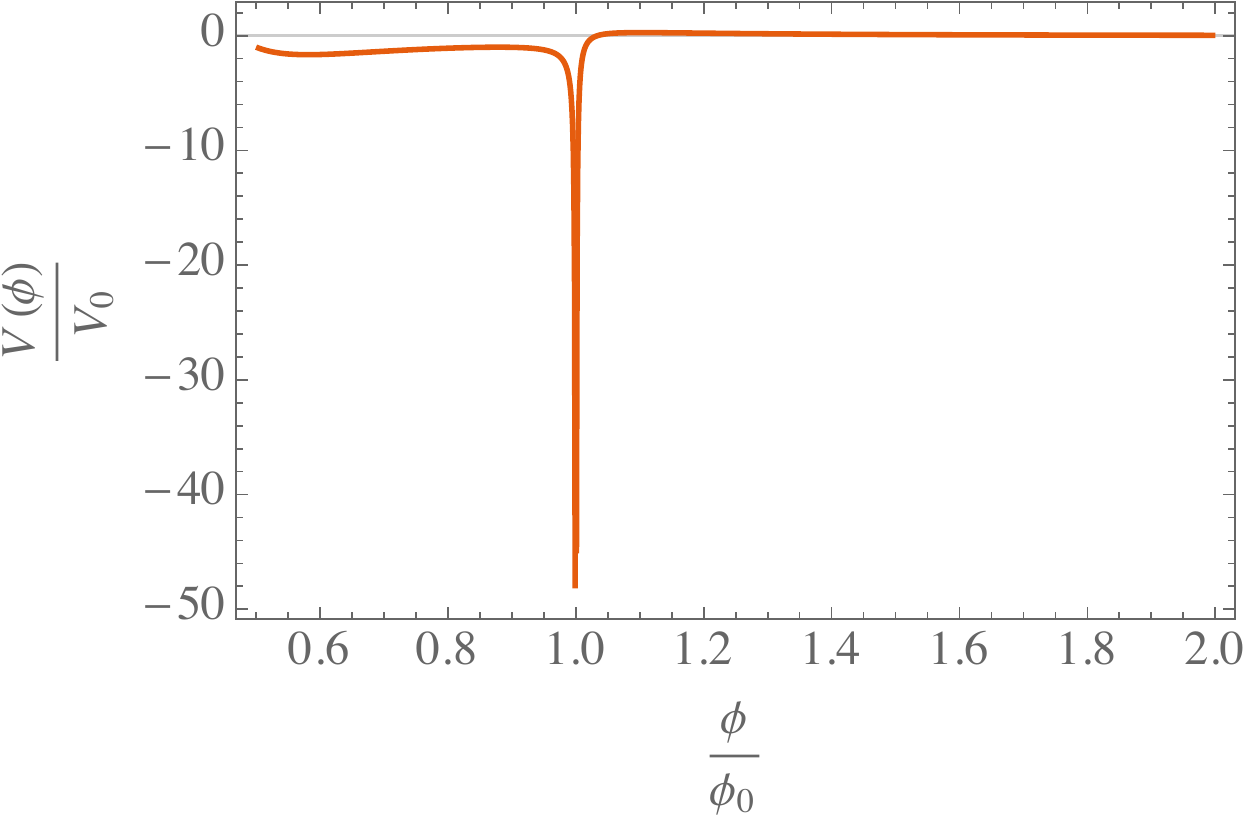}
\includegraphics[width=0.45\textwidth]{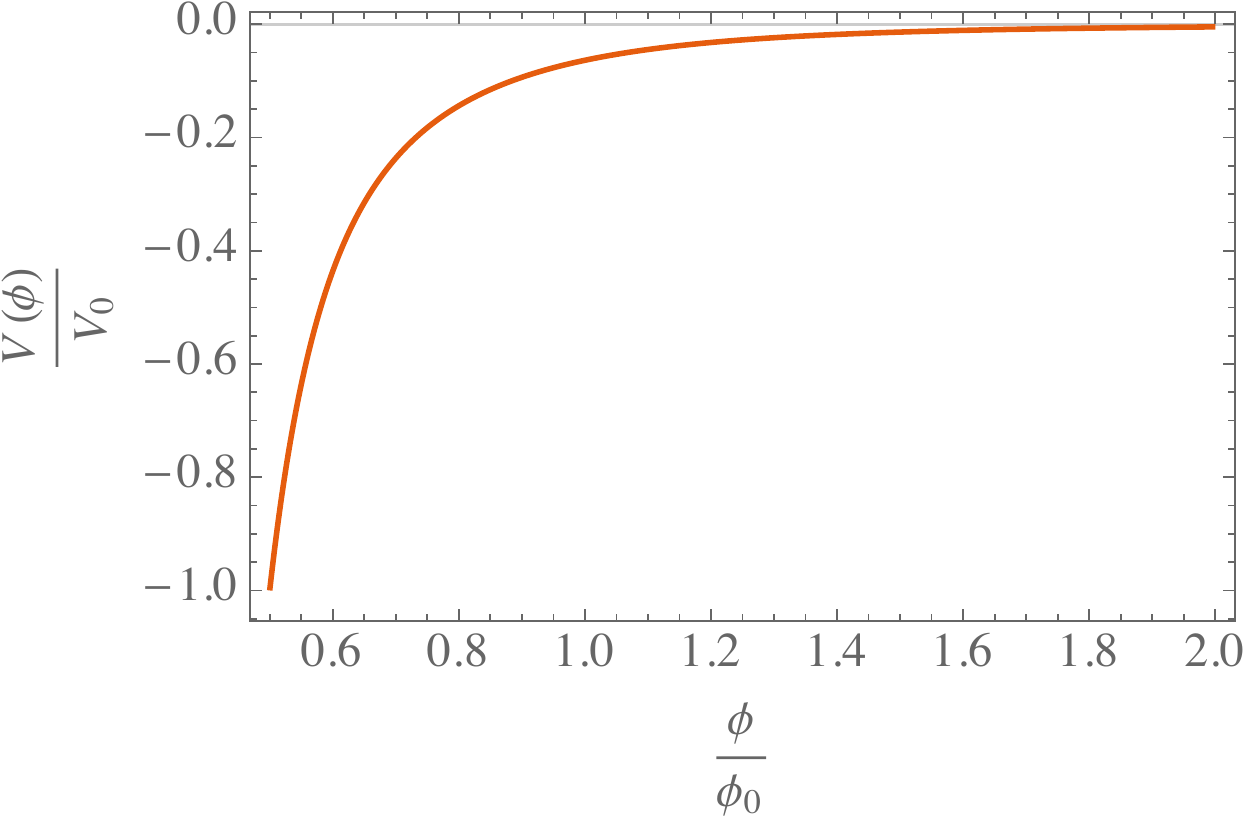}
\caption{The left panel takes the values $\zeta=-1$, $\phi_0=1$, $\phi_1=1$, $\kappa=1$, and 
$\rho_0=1$; and the right panel $\zeta=-1$, $\phi_0=-1$, $\phi_1=1$,
$\kappa=1$, and $\rho_0=1$. 
\label{fig:fig3}}
\end{center}
\end{figure}
In Fig.(\ref{fig:fig3}) it is shown the case where $\zeta =-1$. The left panel shows a nearly flat 
potential in the entire range of $\phi$ with a similar sharp deep well around $\phi=1$ as we see in 
the previous case. The right panel is completely different compared to the previous cases. It is positive 
and decaying in all the range of the scalar field, and diverges at the origin $\phi=0$.

The term $\beta$ in Eq.(\ref{eq:Sol_phi}) will appear constantly from here on in the solution for
stealth and there is no reason to constrain its value, so it could be positive,
imaginary or null, depending if its value becomes greater than, less than or equal
to $\zeta=12/73$. Now the solution (\ref{eq:Sol_phi}) remain valid to the ranges
$\Re-\{(0,12/73)\}-\{\zeta=12/73\}-\{\zeta=1/4\}$. The critical value
$\zeta_c=12/73$, the range $\zeta\in (0,12/73)$ and $\zeta=1/4$ will be analyzed in
separate subsections. It is interesting to note that for values of $\zeta<0$, the
discriminant of the root in $\beta$ always become positive so for this case the
solution given by Eq. (\ref{eq:Sol_phi}) remain valid.

\subsection{Stealth cosmology for $\zeta=\zeta_c$}

\begin{figure}[!]
  \begin{center}   \includegraphics[width=0.45\textwidth]{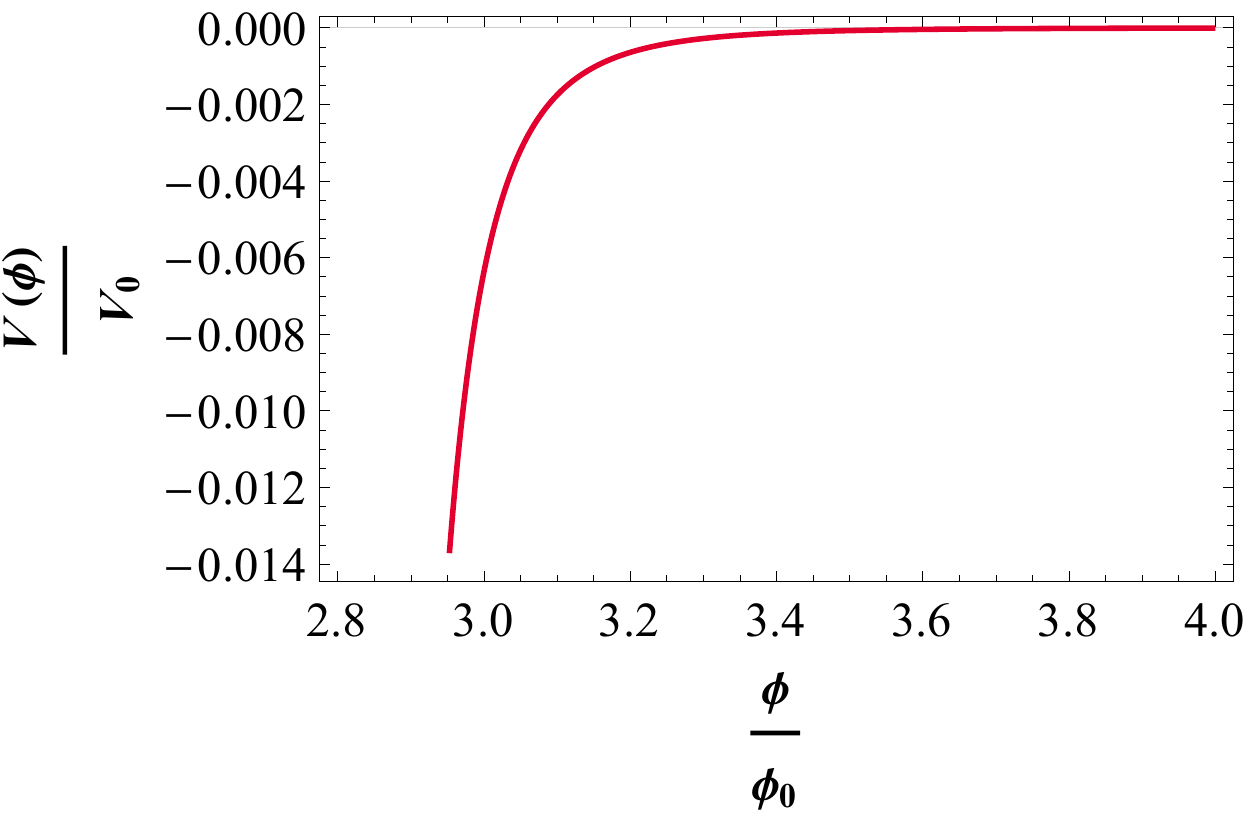}
   \includegraphics[width=0.45\textwidth]{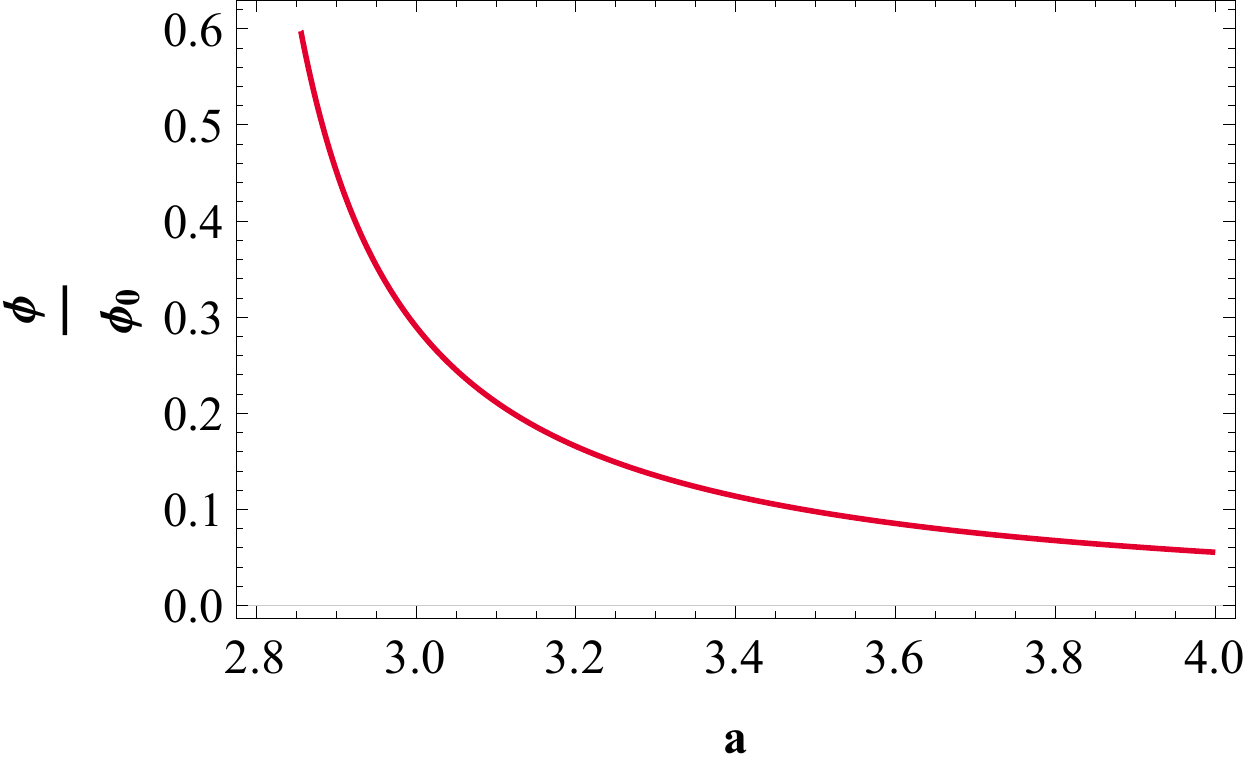}
   
    \caption{At the left we show the self-interaction potential vs.
the scalar field $\phi$ and the graph on the right is for the scalar field vs. is the scale factor. The 
values used are $\zeta=12/73$, $b=1$, $\phi_0=1$, $\phi_1=1$,
$\kappa=1$, and $\rho_0=1$.
} 
\label{fig:ZetasCrit}
  \end{center}
\end{figure}

As was pointed out in the previous section, the value for $\zeta_c=12/73$ is not
described by (\ref{eq:Sol_phi}). In order to get the solution for this value we
substitute directly $\zeta=\zeta_c$ on (\ref{eq:F}), so we have
\begin{eqnarray}\label{eq:phiCrit}
  \phi(a)=(\phi_1 \ln a-\phi_0)^{-\frac{24}{25} }
  a^{-\frac{6}{5}},
\end{eqnarray}
and the potential associated to the above equation becomes
\begin{eqnarray}
  V(\phi)=-\frac{2}{625 }\frac{\kappa\rho_0}{ \sqrt{a}}
  [V_2\ln^2 a   +V_1 \ln a   +V_0]\phi^{\frac{49}{12}}
\end{eqnarray}
where we have defined $V_2:=25(19\zeta-3)\phi_1^2$, $V_1:=50(3-19\zeta)\phi_0\phi_1+20
(43\zeta-6)\phi_1^2$ and $V_0:=25(19\zeta-3)\phi_0^2-20(43\zeta-6)\phi_1\phi_0+
8(73\zeta-6)\phi_1^2$. Once again was verified the right functional form of $\rho$ by substituting 
(\ref{eq:phiCrit}) into (\ref{eq:rho}).
In the figure (\ref{fig:ZetasCrit}) we show the potential to different values of the
parameters, as one can see we do not include the graph for $\phi_1=\phi_0=-1$ because it shows a 
similar as the first graph.

An interesting fact that emerges from the analysis of this solution is that the field $\phi$ only reaches 
real values after a certain time of evolution of the scale factor. So, in this case, the stealth appears 
after of space-time, and the values of $\phi_0$ and $\phi_1$ make the graph start closer or farther 
from the origin.

\subsection{Stealth cosmology for $0<\zeta<\zeta_c$}


From (\ref{eq:stealtuv}) in the range  $0<\zeta<\zeta_c$ the discriminant of $\beta$
is negative, the exponent is complex, so that we have oscillating solutions given by
\begin{eqnarray}
\phi(a)=\Phi_0[a^{5/2}(\phi_1 \sin \gamma - \phi_2 \cos \gamma)]^{\frac{2\xi}{4\xi-1}} 
\qquad \mathrm{where}\qquad
\gamma=\frac{1}{4}\frac{\sqrt{12-73\zeta}}{\sqrt{\zeta}}\ln a \quad
\end{eqnarray}
where $\phi_1$ and $\phi_2$ are constants.  Now, if we define the constants as 
$\phi_1=\sin\delta$ and $\phi_2=\cos\delta$, we can get a simpler scalar field, this is
\begin{eqnarray}
\phi(a)=[a^{\frac{5}{4} }(\cos(\beta\ln a+\delta))
  ]^{\frac{2\zeta}{4\zeta-1}},\quad \mathrm{where}\quad
 \beta:=\frac{1}{4}\sqrt{\frac{12-73\zeta}{\zeta}} 
\end{eqnarray}
and its potential
\begin{eqnarray}
 V(\phi)=\frac{\zeta}{24(4\zeta-1)^2}\frac{\kappa\rho_0}{a^3}\{V_1\phi^2
 -[V_2 a^{\frac{5}{4}}\sin(\beta\ln a+\delta)+V_3 a^{\frac{5}{2}}
  \phi^{\frac{1-4\zeta}{\zeta}}]\phi^{\frac{1}{2\zeta}}\},
\end{eqnarray}
\begin{eqnarray}
V_1&:=&(432\zeta^2-107\zeta + 6),\quad
V_2\,:=\,2(7-48\zeta)\sqrt{\zeta(12-73\zeta)}\quad \mathrm{and} \quad \nonumber\\
V_3&:=&(12-73\zeta).\nonumber
\end{eqnarray}

\begin{figure}[h!]
 \begin{center} 
 \includegraphics[width=0.45\textwidth]{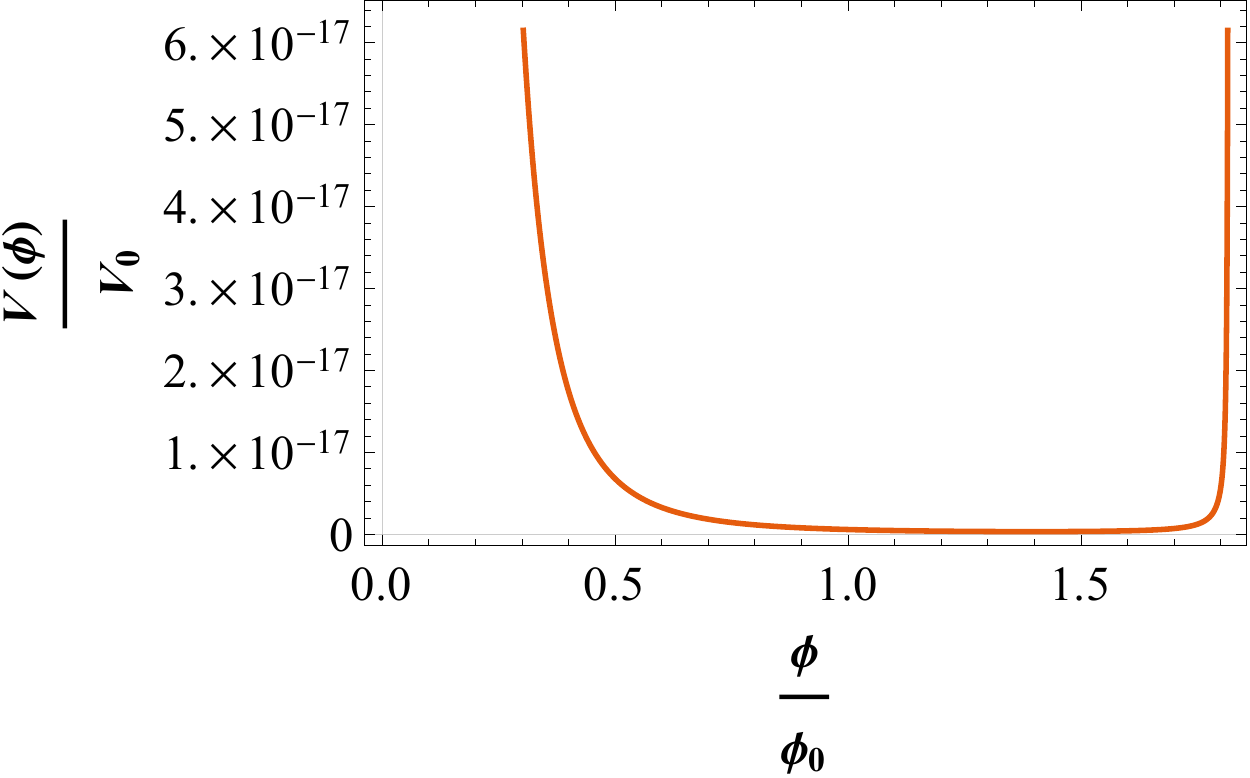}
\includegraphics[width=0.45\textwidth]{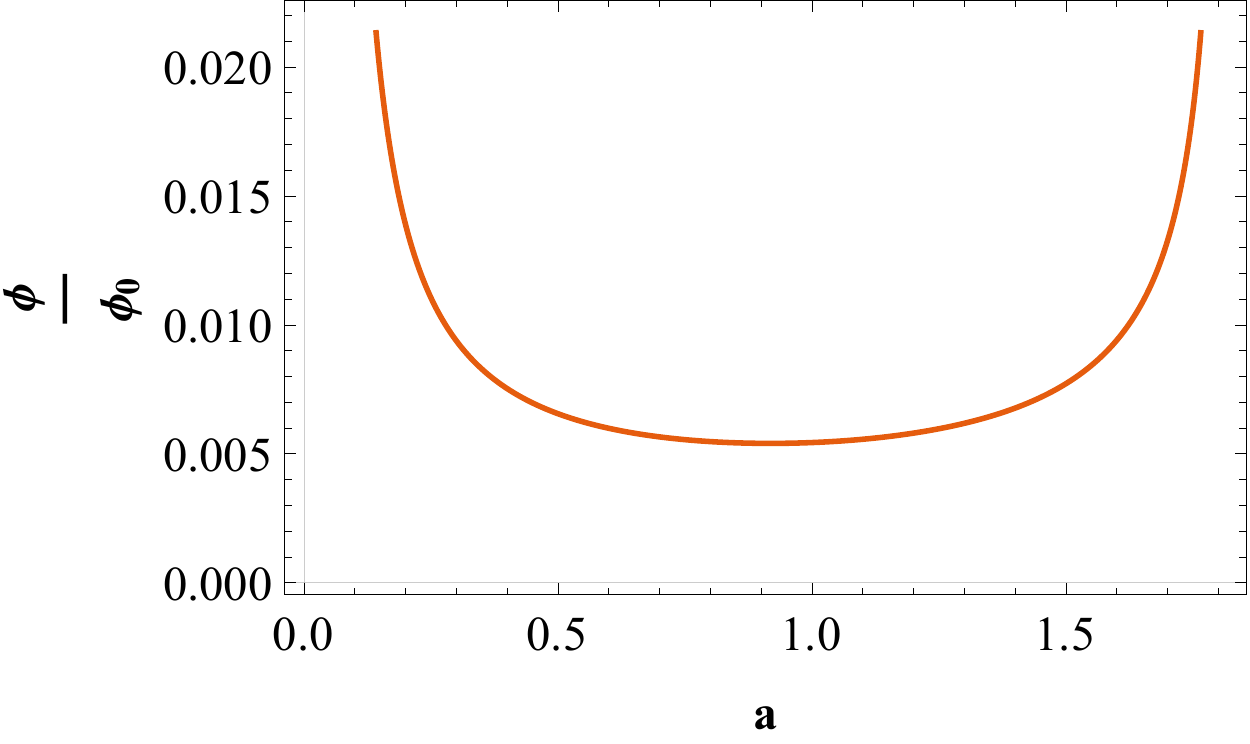}
    \caption{At the left we show the self-interaction potential vs.
the scalar field $\phi$ and
the graph on the right is for the scalar field vs. is the scale factor. The values used are 
$\zeta=10/73$, $\phi_0=1$, $\phi_1=-1$, $\kappa=1$, and $\rho_0=1$.}
\label{ZetasMC}
 \end{center}
\end{figure}

The graph for this case is shown in figure (\ref{ZetasMC}), the stealth field, in this case, have a periods 
of real and complex values, which are longer as the value of the phase into the cosine increase.

\subsection{Stealth cosmologies for $\zeta=1/4$}
\label{sec:2}
\begin{figure}[h!]
  \begin{center}
\includegraphics[width=0.45\textwidth]{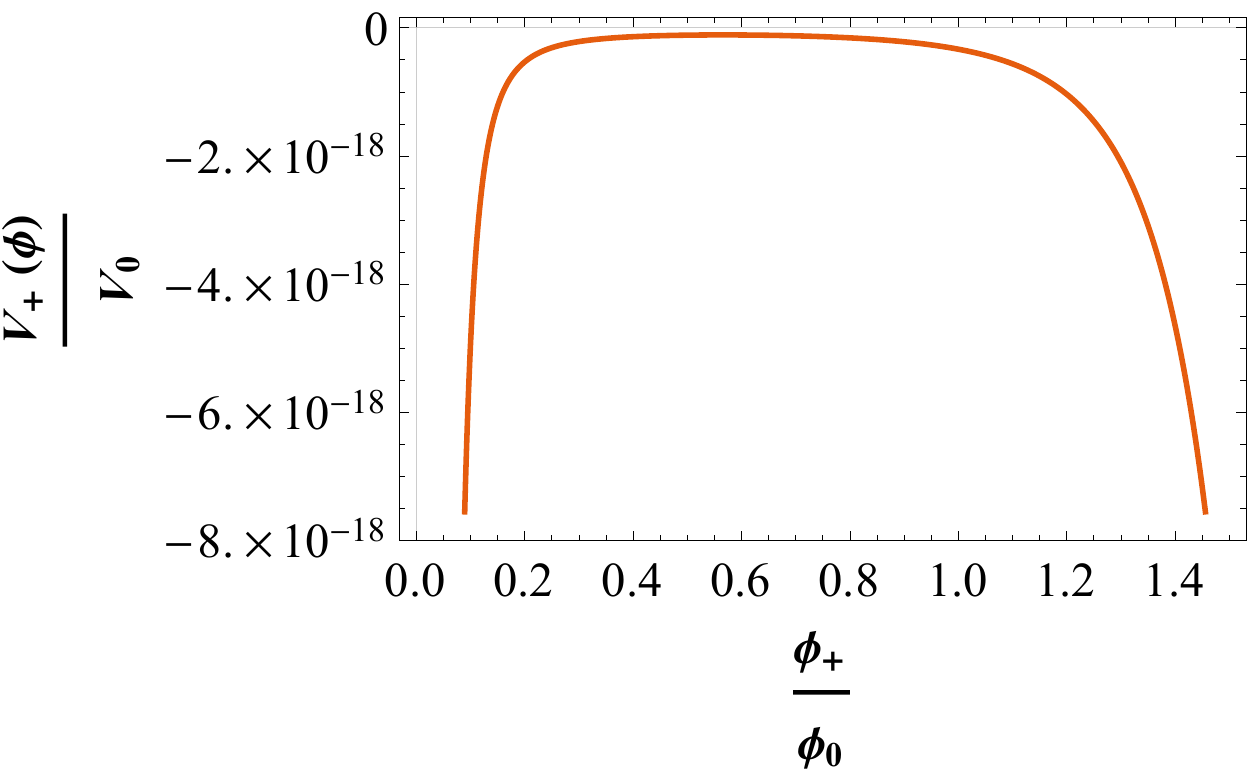}
\includegraphics[width=0.45\textwidth]{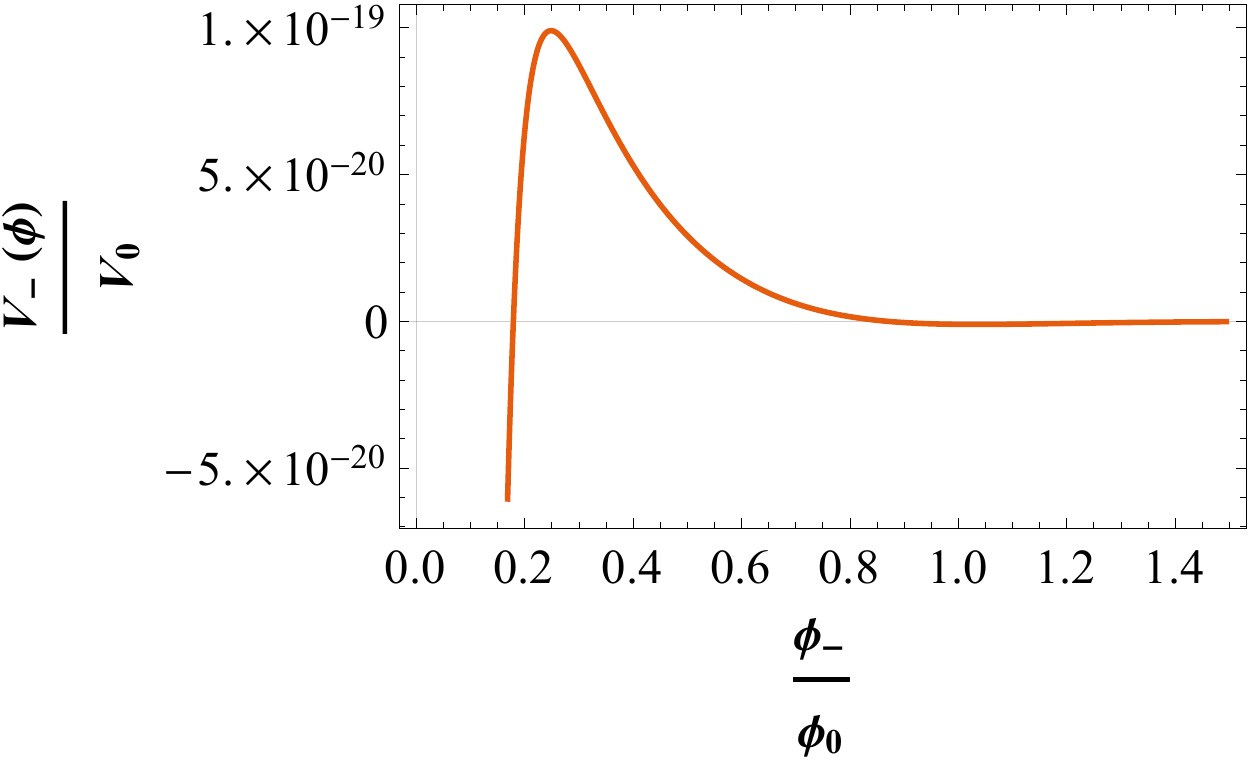}
\caption{At the left we show  $V_{+}$ the self-interaction of the scalar field
vs. $\phi$
for the values $\zeta=1/4$, $\phi_0=1$, $\phi_1=1$, $\kappa=1$, and $\rho_0=1$;
While at the right we show $V_{-}$ vs. $\phi$
for the values $\zeta=1/4$, $\phi_0=1$, $\phi_1=1$, $\kappa=1$, and $\rho_0=1$.}
\label{fig:fig5}
\end{center}
\end{figure}

There is a special case for $\zeta=1/4$.
In particular here it is possible to integrate the general case. Under the change
$\dot{\phi}/\phi:=\dot{f}$ Eq. (\ref{eq:F}) is written as
\begin{eqnarray}
  \frac{d}{dt}f=-\frac{4\zeta-1}{2\zeta}f^2+\frac{\dot{a}}{a}f+\frac{1}{2}
  \frac{\kappa \rho_0}{a^3}.
\end{eqnarray}
Now from the above equation it is evident why this value for $\zeta$ is a special
value, and in general its integral is
\begin{eqnarray}
  \phi(a)_\pm=\phi_1e^{\pm \phi_0 a^{5/2}} a^{-\frac{3}{5}}
\end{eqnarray}
and its self-interaction potential gives
\begin{eqnarray}
  V(a)_{\pm } =- \rho_0 \kappa \left(\frac{25}{24}a^5\phi_0^2
  \pm \frac{3}{4}\phi_0 a^{5/2} + \frac{1}{100} \right) \frac{\phi_{\pm}^2}{a^3}
\end{eqnarray}
The behavior for both potentials is displayed in Fig.(\ref{fig:fig5}).
In the figures of this section we wish study the sensibility of the potential to
the sign of the integration constants. For this case the solution allow stealth at same time as the start 
of spacetime. 
\section{Homogeneous power law cosmologies}

In this section, we show the advantages of our approach by studying the contributions of the stealth 
field to the principal quantities setting up the cosmology from the Einstein equations, that is, the 
energy density and the pressure. In order to prove our approach, we reproduce some of the results 
reported in \cite{Ayon-Beato:2015bsa}, we study the power law $a(t)=a_0 t^b$ for homogeneous 
cosmologies. As in previous case, the behavior of stealth $\phi$
depends on value of $\zeta$ so we study the solutions for the more general cases
reported in the work aforementioned. Even more, in what follows we use the
conformal metric as in \cite{Ayon-Beato:2015bsa},
\begin{eqnarray}\label{conformalM}
ds^2=a(t)^2(-d\tau^2+\frac{dr^2}{1-kr^2}+r^2(d\theta^2+\sin^2\theta d\phi^2))\,.
\end{eqnarray}

In this case from Eq. (\ref{eq:perrona}), we get the expressions for $\rho$ and $p$ 
\begin{eqnarray}
  \rho(t)=-\frac{1}{\kappa}\left[ 6\frac{\dot{\phi}}{\phi}\frac{\dot{a}}{a^3}
  +\frac{1}{2\zeta a^2}\left(\frac{\dot{\phi}}{\phi}\right)^2 +\frac{V(\phi)}
  {\zeta\phi^2}\right],\label{eq:rho2}\\
  p(t)=\frac{1}{\kappa}\left[\frac{2}{a^2}\frac{\ddot{\phi}}{\phi}+\frac{2}{a^2}
  \frac{\dot{\phi}}{\phi}\frac{\dot{a}}{a}+\frac{4\zeta-1}{2\zeta}\frac{1}{a^2}
\left(\frac{\dot{\phi}}{\phi}\right)^2+\frac{V(\phi)}{\zeta\phi^2}\right].
\label{eq:p2} 
\end{eqnarray}
Adding (\ref{eq:rho2}) and (\ref{eq:p2}) and doing the same for the expressions of $p$ and $\rho$ 
obtained from field equations of GR, we obtain the equations for the stealth field and its 
self-interaction potential $V(\phi)$,

\begin{eqnarray}\label{pfs}
\frac{d\ln \phi}{dt}\cdot \frac{d}{dt}\ln\left(\frac{\dot{\phi}^2 \phi^{\frac{2\zeta-1}{\zeta}}}{a^4}\right)
+\frac{d\ln a}{dt}\cdot \frac{d}{dt}\ln\left( \frac{\dot{a}^2}{a^{4}}\right)
=0,
\end{eqnarray}
now subtracting Eqs. (\ref{eq:rho2}) and (\ref{eq:p2}) and by using again $p$ and $\rho$ from GR,
\begin{eqnarray}\label{Vc}
  V(\phi)= -\zeta \left[
              \frac{d \ln \phi}{dt}\cdot\frac{d}{dt} \ln\left( a^4 \phi \dot{\phi}  \right)
              +\frac{d\ln a}{dt}\cdot \frac{d\ln(a\dot{a})}{dt}
              \right]\left(\frac{\phi}{a}\right)^2.
\end{eqnarray}
By substituting (\ref{Vc}) into Eqs. (\ref{eq:rho2}) and (\ref{eq:p2}), we get
\begin{eqnarray}
\kappa a^2  \rho = \frac{d\ln \phi}{dt}\cdot \frac{d}{dt}
         \ln\left(\frac{ \dot{\phi}\phi^{\frac{2\zeta-1}{2\zeta}}}{a^2} \right)
         +\frac{d\ln a}{dt}\cdot \frac{d}{dt}\ln ( a \dot{a})
\end{eqnarray}

\begin{eqnarray}
  \kappa a^2 p =\frac{d \ln \phi}{dt} \cdot \frac{d}{dt}
       \ln\left(\frac{\phi^{\frac{2\zeta-1}{\zeta}}\dot{\phi}}{a^2} \right)
       -\frac{d\ln a}{dt} \cdot \frac{d \ln(a\dot{a})}{dt}
\end{eqnarray}
From the above equations, it is clear that is possible to quantify the contributions to the stealth field 
to $\rho$ and $p$. In order to realize the cosmology the Friedmann equation in the spacetime 
described by (\ref{conformalM}) is $H^2=\kappa a^2 \rho/3$ with state equation 
$p=\omega \rho$, explicitly for Friedmann equation in terms of stealth
\begin{equation}
H^2=\frac{1}{3}\left[ \frac{d\ln \phi}{dt}\cdot \frac{d}{dt}
         \ln\left(\frac{ \dot{\phi}\phi^{\frac{2\zeta-1}{2\zeta}}}{a^2} \right)
         +\frac{d\ln a}{dt}\cdot \frac{d}{dt}\ln ( a \dot{a})\right],
\end{equation}
and for power-law cosmologies, the state equation remains valid. 


\subsection{Homogeneous cosmologies with $\zeta>\zeta_c$}

As in the previous section, the behavior of the solution of (\ref{pfs}) for $\phi$ depends on the value 
of $\zeta$ so we study the solutions in its different regimes or ranges.

\begin{eqnarray}\label{phipf1}
  \phi(t)=\Phi_0 \left[ \left(\frac{t}{ t_0} \right)^{b+\frac{1}{2}+\beta }
  -\left(\frac{t}{ t_0} \right)^{b+\frac{1}{2}-\beta }
  \right]^{\frac{2\zeta}{4\zeta-1} }  \mbox{where}  \quad t_0=\frac{\phi_0}{\phi_1}
\end{eqnarray}
\begin{eqnarray}
  \Phi_0=2^{\frac{2\zeta}{4\zeta-1}}\left[\frac{(4\zeta-1)^2
  (\phi_0\phi_1)^{2\beta}}{4\zeta^3\beta^2}\right]^{\frac{\zeta}{4\zeta-1}}\quad;
  \beta=\frac{\sqrt{\zeta(12b^2+12b+1)-2(b+1)b}}{2\zeta}; \quad
  \nonumber
\end{eqnarray}
A contribution of our work to the solutions found in \cite{Ayon-Beato:2015bsa} are the expressions 
for the energy density and pressure given by the Eqs. (\ref{eq:rho2}) and (\ref{eq:p2}), respectively, 
\begin{eqnarray}
 \kappa a^2\rho &=& \frac{\ddot{a}}{a}
       +\left(\frac{\dot{a}}{a}\right)^2
       +4\frac{\beta\zeta}{4\zeta-1}\left[\frac{b}{t^2}
       - \frac{\dot{a}}{a}\frac{1}{t}
       \right]\frac{\phi_{+}}{\phi_{-}}
         -2(2b+1) \frac{\zeta}{4\zeta-1}\frac{\dot{a}}{a}\frac{1}{t}
         \nonumber\\ &{}&
          +\frac{\zeta}{2(4\zeta-1)}\frac{(4b^2+4\beta^2-1)}{t^2}
\end{eqnarray}

\begin{eqnarray}
 \kappa a^2 p &=&-\frac{\ddot{a}}{a}
       -\left(\frac{\dot{a}}{a}\right)^2
       +4\frac{\beta\zeta}{4\zeta-1}\left[\frac{b}{t^2}
       - \frac{\dot{a}}{a}\frac{1}{t}
       \right]\frac{\phi_{+}}{\phi_{-}}
         -2(2b+1) \frac{\zeta}{4\zeta-1}\frac{\dot{a}}{a}\frac{1}{t}
         \nonumber\\ &{}&
          +\frac{\zeta}{2(4\zeta-1)}\frac{(4b^2+4\beta^2-1)}{t^2}
\end{eqnarray}

\begin{eqnarray}
  \phi_{\pm}= \left[ \left(\frac{t}{ t_0} \right)^{b+\frac{1}{2}+\beta }
  \pm \left(\frac{t}{ t_0} \right)^{b+\frac{1}{2}-\beta }
  \right]
\end{eqnarray}
It is worth mentioning that all terms in the above expressions which depend explicitly on time are 
contributions of the stealth field. The term $\phi_{+}/\phi_{-}$ vanish with the contribution of 
$\dot{a}/a$. 

\subsection{Homogeneous cosmologies for $\zeta=\zeta_c$}

At the value of $\zeta=\zeta_c$ Eq.(\ref{phipf1}) fail so this case must be
integrated in separate way. Substituting this value into Eq.(\ref{pfs}) the stealth
field
\begin{eqnarray}
  \phi(t)=\Phi_0\Big[ t^{\frac{1}{2}(2b+1) }
  (\phi_1\ln t-\phi_0)\Big]^{\frac{-4b(b+1)}{(2b+1)^2} };\quad \mbox{where}\quad
  \Phi_0:=\left(\frac{4b(b+1)}{(2b+1)^2}\right)^{\frac{-4b(b+1)}{(2b+1)^2} },
\end{eqnarray}
here $\phi_1$ and $\phi_0$, are integration constants.
For this ranges the energy-density and the pressure are given by

\begin{equation}
 \kappa a^2\rho = \frac{\ddot{a}}{a}+\left(\frac{\dot{a}}{a}\right)^2+
 8\frac{b(b+1)}{2b+1}
 \left[\frac{b}{t^2}-\frac{\dot{a}}{a}\frac{1}{t}\right]\frac{\phi_1}{\phi_{-}}
 -4\frac{b(b+1)}{2b+1}\frac{\dot{a}}{a}\frac{1}{t} +\frac{b(b+1)(b-1)}{2b+1}\frac{1}
 {t^2},
\end{equation}

\begin{equation}
\kappa a^2 p =-\frac{\ddot{a}}{a}-\left(\frac{\dot{a}}{a}\right)^2+
 8\frac{b(b+1)}{2b+1}
 \left[\frac{b}{t^2}-\frac{\dot{a}}{a}\frac{1}{t}\right]\frac{\phi_1}{\phi_{-}}
 -4\frac{b(b+1)}{2b+1}\frac{\dot{a}}{a}\frac{1}{t} +\frac{b(b+1)(b-1)}{2b+1}\frac{1}
 {t^2},
\end{equation}
where we have defined $\phi_{\_}:=\phi_1 \ln t-\phi_0$.

\subsection{Homogeneous cosmologies for $\beta<0$}
There is a range for $\zeta$ and $b$ where the discriminant of the solution becomes to be negative; 
for this range the solutions are oscillating. At a difference of the reported in 
\cite{Ayon-Beato:2015bsa}, we do not use the relic symmetries approach. So we present directly the 
solution of (\ref{pfs}) and shown the expressions for $\rho$ and $p$. In both cases our approach 
work. 
 \begin{eqnarray}
   \phi(t)=\Phi_0(t^{\frac{2b+1}{2} }[\phi_1\sin(\beta\ln t)
   +\phi_0\cos(\beta\ln t)])^\frac{2\zeta}{4\zeta-1},
\end{eqnarray}
\begin{eqnarray}
  \Phi_0:&=&\left(-\frac{(4\zeta-1)^2}{\zeta(12b^2+12b+1)-2b(b+1)}
   \right)^{\frac{\zeta}{4\zeta-1} },\nonumber\\
   \beta:&=&\frac{1}{2}\sqrt{
   \frac{\zeta(12b^2+12b+1)-2b(b+1)}{\zeta} } \nonumber
 \end{eqnarray}
 $\phi_1$, $\phi_0$ are integration constants.
 Now the energy density and the pressure are
 \begin{eqnarray}
 \kappa a^2\rho&=&\frac{\ddot{a}}{a}+\left(\frac{\dot{a}}{a}\right)^2+
 \frac{4\zeta\beta}{4\zeta-1}\left(\frac{\dot{a}}{a}\frac{1}{t}-
 \frac{b}{t^2}\right)\frac{\phi_{-}}{\phi_{+}}-\frac{2\zeta(b+1)}{4\zeta-1}
 \frac{\dot{a}}{a}\frac{1}{t}\nonumber\\ &{}&
 -\frac{(4\beta^2-4b^2+1)\zeta}{2(4\zeta-1)}
 \frac{1}{t^2},
\end{eqnarray}

\begin{eqnarray}
 \kappa a^2 p&=&-\frac{\ddot{a}}{a}-\left(\frac{\dot{a}}{a}\right)^2+
 \frac{4\zeta\beta}{4\zeta-1}\left(\frac{\dot{a}}{a}\frac{1}{t}-
 \frac{b}{t^2}\right)\frac{\phi_{-}}{\phi_{+}}-\frac{2\zeta(b+1)}{4\zeta-1}
 \frac{\dot{a}}{a}\frac{1}{t} \nonumber\\ &{}&
 -\frac{(4\beta^2-4b^2+1)\zeta}{2(4\zeta-1)}
 \frac{1}{t^2},
\end{eqnarray}
we have defined $\phi_{\pm}:=\phi_1\sin(\beta\ln t)\pm \phi_0\cos(\beta\ln t)$,
note that vanish linear contributions in $\beta$ as in first case, and it is
necessary to recovery the behavior of $\rho$ and $p$.

Finally we recover the state equation $p=\omega \rho$ since all the contributions
to density energy and pressure are $\propto t^{-2(b+1)}$ which are the predict
behavior of GR, at difference in this theory, we quantify the contributions from
stealth field.

We decided don't include the potentials for all the cases reviewed here cause has
been studied very well in \cite{Ayon-Beato:2015bsa}, and the purpose of this
work it is to shown the quantify relations of stealth contributions to the
cosmology.



\section{Conclusions}

%
The main goal of this work is to provide a novel approach to construct cosmological models endowed 
with stealth. We studied the stealth configurations in presence of sources, choosing a perfect fluid 
and particularizing it to study two cases, dust, and the power law, both in a homogeneous 
cosmology. As a result, we obtain a general approach to relate the stealth with any kind of matter 
coupled to the gravitational field whenever there is stealth on it. As the stealths are such that they do 
not warp the background spacetime, intuitively expected to play a role in the cosmological dynamics. 
In this sense, there was provided a way to be quantified their contributions, namely, through 
expressions for energy density and the pressure. This was achieved by giving the matter information 
to the system of equations which gives room to a stealth field configuration instead of the geometrical. 

For completeness, we show the behavior of some cases of the self-interaction potential for the 
stealth, focusing on the sensibility of the sign of the integration constants. The reason for this last 
issue because we expect that the parameter $\zeta$ be fixed when the model will be contrasted 
against observational data as was shown in \cite{1972}. From the analysis of the solutions, the 
stealth is allowed from the beginning of spacetime, except when $\zeta=\zeta_c$ where its 
existence starting after the beginning of spacetime.  

In future works, we will use the approach to study subclasses of the Hordensky's theories which 
have self-tuning mechanism, for example, those reported in \cite{4fabs}. Additionally, we explore 
the perturbations like in \cite{odinstov}, where some clues that what could happen with our 
proposal.

\section{Acknowledgments}

Authors thank to Eloy Ay\'on-Beato for enlightening discussions.
 Special thanks to Carlos Manuel Rodr\'{i}guez.
CC acknowledges partial support by CONACyT Grant CB-2012-177519-F
and grant PROMEP, CA-UV, \'Algebra, Geometr\'{\i}a y
Gravitaci\'on.  This work was partially supported by SNI (M\'exico).
VHC acknowledges partial support by DIUV-REG-50/2013.
AA acknowledges partial support by CONACyT Grant Estancias Posdoctorales
Vinculadas al Fortalecimiento de Calidad del Posgrado Nacional 2016.

\section{Appendix A: Stealth field equation}\label{app:A}
We study the consequences of the stealth in cosmology,
Einstein’s field equations for a perfect fluid source are furnished by
\begin{eqnarray}
G_{\mu\nu}-\kappa T^m_{\mu\nu} =0=\kappa T^s_{\mu\nu}
\end{eqnarray}
here $G_{\mu\nu}$ is are the Einstein tensor, $T^m_{\mu\nu}$ is the energy momentum tensor of a 
perfect fluid
\begin{eqnarray}
T_{\mu\nu}^m=(\rho+p)u_\mu u_\nu +pg_{\mu\nu}\,,
\end{eqnarray}
where $p=p(t)$ is the pressure, $\rho=\rho(t)$ the energy density, $u_{\mu}$is the fluid velocity four 
vector with $u_\mu u^\mu=1$. $T^s_{\mu\nu}$ is the stress energy tensor of the stealth 
field (\ref{eq:stealtuv}).

The Friedmann-Lemaitre-Robertson-Walker metric for a homogeneous isotropic universe is provided 
by
\begin{eqnarray}
 ds^2=-dt^2+\frac{a^2}{1-kr^2}dr^2+a^2(r^2d\theta^2+r^2 sin(\theta)^2 d\phi^2)
\end{eqnarray}
where $k=\pm 1,0$,  indicating the spatial curvature constants. In the case $k=0$ the field equations 
for the background 
are given by
\begin{eqnarray}
\frac{\dot{a}^2}{a^2}=\frac{\kappa}{3}\,\rho(t)\,,\label{eq:11}
\\
2\frac{\ddot{a}}{a}+\frac{\dot{a}^2}{a^2}=-\kappa \,p(t),\label{eq:12}
\end{eqnarray}
and the continuity equation that satisfies the fluid is
\begin{eqnarray}
\dot{\rho} +3H(\rho+p)=0\,.\label{eq:13}
\end{eqnarray}
Now, we assume that the universe is spatially flat and the dust case, i.e. when the pressure is zero. 
We go back to the equations (\ref{eq:11}) and (\ref{eq:13}), to obtain that density and  scale factor 
are given explicitly in the form
\begin{eqnarray}
\rho(t)&=&\rho_0/a^3\,,\\
a(t)&=&(a_1t+a_0)^{2/3}\quad \mathrm{where}\quad a_1^2=\kappa\rho_0/3\,\label{eq:02}\,.
\end{eqnarray}
in thats case the equations for the stealth are given by the Eq.(\ref{eq:rho}). 

Expanding Eq.(\ref{eq:rho}), explicitly we get
\begin{eqnarray}
\kappa \rho=\kappa\frac{\rho_0}{a^3}=2 \frac{\ddot{\phi}}{\phi} 
   +\frac{(2\zeta-1)}{\zeta}
  \left(\frac{\dot{\phi}}{\phi}\right)^2-2\frac{\dot{\phi}}{\phi 
  }\frac{\dot{a}}{a},
\end{eqnarray}

now, in order to get an analytical solution  of the stealth field, we propose $\phi=\phi(a(t))=\phi(a)$ in
the Eq. (\ref{eq:rho}).  Now, the field equation is given by
 \begin{eqnarray}
  \frac{1}{\phi}\left(\frac{d^2\phi}{da^2}
  \dot{a}^2+\frac{d\phi}{da}\ddot{a} \right)
  +\frac{(2\zeta-1)}{2\zeta}\frac{1}{\phi^2}\dot{a}^2
  \left(\frac{d\phi}{da}\right)^2-\frac{1}{\phi a}\frac{d\phi}{da}\dot{a}^2
  =\frac{\kappa\rho_0}{2 a^3}\label{eq:16}
\end{eqnarray}
From (\ref{eq:02}) is easy to shown
\begin{eqnarray}
\dot{a}^2=\frac{a_1^2}{a}\quad \mathrm{and} \quad \ddot{a}=-\frac{\dot{a}^2}{2 a},
\end{eqnarray}
then equation (\ref{eq:16}) finally have the form
\begin{eqnarray}
  \left[\frac{1}{\phi}\frac{d^2\phi}{da^2}
 -\frac{3}{2 a}\frac{1}{\phi}\frac{d\phi}{da}
  +\frac{(2\zeta-1)}{2\zeta}\frac{1}{\phi^2}
  \left(\frac{d\phi}{da}\right)^2\right]\dot{a}^2
  =\frac{\kappa\rho_0}{2 a^3}\label{eq19}.
\end{eqnarray}
factorizing $\frac{1}{\phi}\frac{d\phi}{da}=\frac{d \ln\phi}{da}$ and rearranging terms and by 
substituting the expressions of $\dot{a}^2$ and $a_1^2$ one arrive at Eq.(\ref{eq:F}).



\end{document}